\documentclass[amssymb,11pt]{article}
\usepackage[pdftex,colorlinks=true,urlcolor=black,filecolor=black,linkcolor=black,
            pdftitle={Quantum Imploding Scalar Fields.},
            pdfauthor={Mark D. Roberts},
            pdfsubject={Canonical quantum Gravity},
            pdfkeywords={imploding scalar},
            pagebackref,pdfpagemode=None,bookmarksopen=true]{hyperref}
\usepackage{amssymb}
\usepackage{graphicx}
\newcommand{\bc}{\begin{center}}
\newcommand{\ec}{\end{center}}
\newcommand{\be}{\begin{equation}}
\newcommand{\ee}{\end{equation}}
\newcommand{\ber}{\begin{eqnarray}}
\newcommand{\ear}{\end{eqnarray}}

\begin{document}
\title{Quantum Imploding Scalar Fields.}
\author{
\href{http://www.violinist.com/directory/bio.cfm?member=robemark}
{Mark D. Roberts},\\
}
\date{$12^{th}$ of March 2017}
\maketitle
\begin{abstract}
The d'Alembertian $\Box\phi=0$ has solution $\phi=f(v)/r$,
where $f$ is a function of a null coordinate $v$,
and this allows creation of a divergent singularity out of nothing.
In scalar-Einstein theory a similar situation arises both for the scalar field
and also for curvature invariants such as the Ricci scalar.
Here what happens in canonical quantum gravity is investigated.
Two minispace Hamiltonian systems are set up:
extrapolation and approximation of these indicates that
the quantum mechanical wavefunction can be finite at the origin.
\end{abstract}
\vspace{2cm}
\tableofcontents
\newpage
\section{Introduction}\label{intro}
For centuries physicists have wondered what happens
at the origin of the reciprocal potential $1/r$,
which is ubiquitous and for example occurs in electromagnetism and gravitation.
For a minimal scalar field obeying $\Box\phi=0$ the situation is worse because
\begin{equation}
\label{firsteq}
\phi=\frac{f(v)}{r},
\end{equation}
where $f$ is a suitable differentiable function of the null coordinate $v$,
is a solution allowing creation of a reciprocal singularity out of nothing
at the origin of the coordinates $r=0$.
The easiest way to avoid this problem is to say that minimal scalar theory
breaks down and another theory is applicable to the problem at hand.
There are a huge number of theories to choose from,
for example Born-Infeld theory \cite{borninfeld} was created partially
to avoid singularities at the origin.
In quantum field theory the scalar field is usually quantized directly
so it is hard to compare with the exact solution (\ref{firsteq}).
In general relativity (\ref{firsteq}) was generalized in 1985 \cite{mdr86,mdr89}
to a solution of the scalar-Einstein equations:
one can have solutions with $\phi$ of the same form but then one needs a compensating
null radiation field,  if the null radiation field is taken to vanish then one ends up
with a simple scalar-Einstein solution;
again one has a scalar field singularity at the origin of the coordinates
and there is also a singularity of the spacetime curvature,
and in this sense the situation is worse that (\ref{firsteq})
because spacetime has also broken down.
The scalar-Einstein solution has at least six related applications.
{\it Firstly} to cosmic censorship:  it is known that in most cases static scalar-Einstein
spacetimes do not have event horizons \cite{chase} and the existence of the solution shows
that this is also the case in one particular instance in dynamic spacetimes.
Whether event horizons actually exist is now a matter of astrophysical observation \cite{NM}
and \cite{AKL}.
{\it Secondly} to numerical models of gravitational collapse \cite{choptuik,gundlach} where it
is a critical value between different behaviours.
{\it Thirdly} to quantum field theories on curved spacetimes where the scalar field can be
equated to the field of the quantum field theory:  whether this is an allowable method or not
is undecided,  in any case it turns out that there are many technical problems concerning
whether objects such as the van Vlech determinate converge fast enough.
{\it Fourthly} to the Hawking effect \cite{hawking},
can the exact solution scalar field be equated to scalar fields created in this,
a related paper is \cite{tominatsu},
{\it Fifthly} annihilation and creation operators,
perhaps these in some way correspond to imploding and exploding fields;
usually these are defined on a fixed background however as geometry is related to matter
there must be a simultaneous change in the gravitational field and perhaps a
preferred graviton configuration,
{\it Sixthly} to canonical quantum gravity which is the subject of the present paper.

It is common in physics to let algebraic expressions to become functions:
however there is not an established word to describe this.
When the algebraic expression is just a constant this is sometimes
referred to as letting the object 'run',  but here sometimes the
algebraic expression is a constant times a variable.
The words 'functionify' and 'functionification' do not appear in dictionaries
so here the word 'relax' is used to describe this process.
Section \S\ref{ses} describes the properties of the scalar-Einstein solution needed here,
in particular the original single null and double null forms are presented,
brute force methods applied to these forms leads to two variable problems.
The solution has two characteristic scalars:  the scalar field and the homothetic Killing
potential,  expressing the solution in terms of these leads to one variable problems.
Section \S\ref{relaxk} describes how to get a Hamiltonian and quantize the system when
the homothetic Killing vector is relaxed,  this can be pictured as what happens when
there is one quantum degree of freedom introduced into the system corresponding to
fluctuations in the homothetic Killing vector away from its classical properties,
classical fluctuation have been discussed by Frolov \cite{frolov}.
Section \S\ref{relaxp} describes how to get a Hamiltonian and quantize
when the scalar field is relaxed.
Section \S\ref{eanda} discusses how to fit the results of the previous two sections together
and many of the assumptions of the model.
Section \S\ref{conc} discusses speculative applications and concludes.
Conventions used are signature $-+++$,
indices and arguments of functions left out when the ellipsis is clear,
$V$ to describe a scalar field potential and
$U$ to describe the Wheeler-DeWitt potential,
$\phi$ for the scalar field in a scalar-Einstein solution,
$\xi$ for a source scalar field,
field equations ${\mathcal G}_{\mu\nu}=G_{\mu\nu}-8\pi\kappa T_{\mu\nu}$
$\mu,\nu,\dots$ are spacetime coordinates,
$A,B,\dots$ are field variables.
\section{The Scalar-Einstein solution}\label{ses}
The solution in the original single null coordinates \cite{mdr86,mdr89,mdrhi} is
\begin{eqnarray}
\label{original}
ds^2&=&-(1+2\sigma)dv^2+2dvdr+r(r-2\sigma v)d\Sigma_2^2,\\
d\Sigma_2^2&\equiv&d\theta^2+\sin(\theta)d\phi^2,~~~
\phi=\frac{1}{2}\ln\left(1-\frac{2\sigma v}{r}\right),~~~
\nonumber
\end{eqnarray}
the Ricci scalar is given by
\begin{equation}
\label{rsorig}
R=\frac{2\sigma^2v}{r^2(r-2\sigma v)^2}\left((1+2\sigma)v-2u\right),
\end{equation}
with other curvature invariants such as the Riemann and Weyl tensors squared being
simple functions of it.
The homothetic Killing vector is
\begin{equation}
\label{hkv}
K=Cv\left(2r+(1-2\sigma)v\right),~~~
K_aK^a=-4CK,~~~
K_{;ab}=-2Cg_{ab},
\end{equation}
with conformal factor $-2C$.
Defining the null coordinate
\begin{equation}
\label{defu}
u\equiv (1+2\sigma)v-2r,
\end{equation}
the solution takes the double null form
\begin{eqnarray}
\label{dnform}
ds^2&=&-dudv+r_+r_-d\Sigma_2^2,~~~
d\Sigma_2^2=d\theta^2+\sin(\theta)^2d\phi^2,\\
r_\pm&=&(1\pm2\sigma)v-u,~~~
\phi=\frac{1}{2}\ln\left(\frac{r_-}{r_+}\right),~~~
K=Cuv,~~~
R=\frac{2\sigma uv}{r_+^2r_-^2}
\nonumber
\end{eqnarray}
To transform the line element to a form in which the scalar field and homothetic Killing
potential are coordinates define
\begin{equation}
\label{xycoords}
y\equiv\frac{K}{C}=uv,~~~
v^2=\frac{y}{1+2\sigma f(x)},~~~
u^2=y\left(1+2\sigma f(x)\right),
\end{equation}
$f=\coth$ gives the region $uv>0$
\begin{equation}
\label{slst}
ds^2=-\frac{dy^2}{4y}+\frac{\sigma^2y}{sl(x)^2}dx^2+\frac{\sigma^2y}{sl(x)}d\Sigma_2^2,~~~~~
x=\phi,~~~
R=2g^{xx},
\end{equation}
$f=\tanh$ gives the region $uv<0$
\begin{equation}
\label{clst}
ds^2=-\frac{\sigma^2 y}{cl(x)^2}dx^2+\frac{dy^2}{4y}+\frac{\sigma^2y}{cl(x)}d\Sigma_2^2,
\end{equation}
where the functions $sl$ and $cl$ are given by
\begin{eqnarray}
\label{defslcl}
sl(x)&\equiv&\sinh(x)(\sinh(x)+2\sigma\cosh(x)),\\
cl(x)&\equiv&\cosh(x)(\cosh(x)+2\sigma\sinh(x)),
\nonumber
\end{eqnarray}
with the properties
\begin{eqnarray}
\label{slclprop}
&&cl-sl=1,~~~~~
sl''=4sl+2,~~~~~
cl''=4cl-2,\\
\nonumber
&&sl'=2\sqrt{sl^2+sl+\sigma^2},~~~~~
cl'=2\sqrt{cl^2-cl+\sigma^2},\\
\nonumber
&&sl(\phi)=\frac{4\sigma^2 uv}{r_+r_-}=2\sqrt{2\sigma^3yR},~~~
cl(\phi)=\frac{1}{r_+r_-}(v-u)((1-4\sigma^2)v-u).
\end{eqnarray}

Properties of this solution such as junction conditions have recently been discussed \cite{mdrhi}.
\begin{figure}[!h]
\centering
\includegraphics[height=6cm]{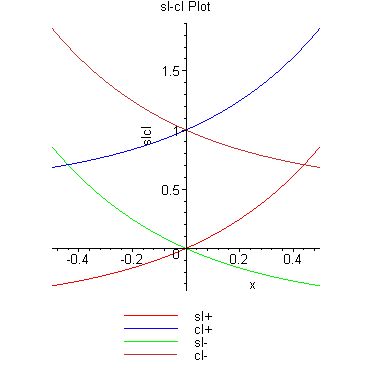}
\caption{sl and cl,  $\sigma=\pm1/2$}
\label{DiagramOne}
\end{figure}
\newpage
\section{Relaxation of the homothetic Killing potential}\label{relaxk}
Consider the line element (\ref{slst}),  let $y=t^2$ then relax $\sigma t$
to become a 'scale factor' function $a(t)$
\begin{equation}
\label{pseudorw}
ds^2=-dt^2+a(t)^2\left[\frac{dx^2}{sl(x)^2}+\frac{d\Sigma^2_2}{sl(x)}\right],
\end{equation}
which is similar to the Robertson-Walker line element the difference being that (\ref{pseudorw})
involves the function $sl(x)$,  defined in (\ref{defslcl}).
Scalar-Einstein Robertson-Walker solutions have been discussed in \cite{mdr23}
and their quantum cosmology in \cite{mdrnq}.
Couple the line element (\ref{pseudorw}) to the source
\begin{equation}
\label{source}
R_{\mu\nu}=2\xi_\mu\xi_\nu+g_{\mu\nu}V_1(\xi),
\end{equation}
to form field equations ${\mathcal G}^{ab}$ which is the Einstein tensor with the source subtracted off.
Having ${\mathcal G}_{x}^t$ necessitates $\xi_t\xi_x=0$ take $\xi_t=0$;
$\xi_x=1$ is forced by the requirement that ${\mathcal G}_{b}^a$ is independent of $x$.
After using the differential properties of $sl$ see (\ref{slclprop}) the field equations become
\begin{equation}
\label{fldeq1}
-a^2{\mathcal G}_{t}^t=3\dot{a}^2-3\sigma^2+a^2V_1,~~~
-a^2{\mathcal G}_{x}^x=-a^2{\mathcal G}_{\theta}^\theta=2a\ddot{a}+\dot{a}^2-\sigma^2+a^2V_1.
\end{equation}
The momentum and Hamiltonian can be read off
\begin{equation}
\label{ham1}
\pi_a=3a\dot{a},~~~~~
{\mathcal H}_1=\frac{\pi_a^2}{6a}+U_1=-\frac{a^3}{2}{\mathcal G}_{t}^t,~~~~~
U_1\equiv-\frac{3}{2}\sigma a+\frac{1}{2}a^3V_1.
\end{equation}
The $q$ Hamiltonian equation is immediate,  the $\pi$ Hamiltonian equation is
\begin{equation}
\label{heq1}
\dot{\pi_a}+\frac{\partial {\mathcal H}_1}{\partial a}=-\frac{3a^2}{2}{\mathcal G}_{x}^x.
\end{equation}
The mini-metric is
\begin{equation}
\label{mini1}
{\mathit M}^{aa}_1=\frac{1}{6a},~~~~~
\det({\mathit M}_1)=6a,
\end{equation}
which has vanishing mini-curvature.
Using the quantization substitution
\begin{equation}
\label{quant}
\pi_A\rightarrow-\imath\hbar\nabla_A
\end{equation}
so that the Hamiltonian (\ref{ham1}) becomes
the Wheeler \cite{wheeler} - DeWitt \cite{dewitt} equation
\begin{equation}
\label{qham1}
{\mathcal H}_1\psi=-\frac{\hbar^2}{6a}\Box\psi+U\psi.
\end{equation}
Using the mini-metric (\ref{mini1})
\begin{equation}
\label{boxpsi1}
\Box\psi=\frac{1}{\sqrt{6a}}\left(\frac{\psi_a}{\sqrt{6a}}\right)_{,a},
\end{equation}
the Hamiltonian (\ref{qham1}) becomes
\begin{equation}
\label{wdw1}
-\frac{72a^3}{\hbar^2}{\mathcal H}_1\psi=2a\psi_{,aa}-\psi_{,a}
+\frac{36a^4}{\hbar^2}\left(3\sigma^2-a^2V_1\right)\psi.
\end{equation}
For $V_1=0$ maple finds a solution that is a linear combination of Bessel functions $B^J_Y$
\begin{equation}
\label{sol1}
\psi_1=\sum_{JY}C^J_Ya^\frac{3}{4}B^J_Y
\left(\frac{3}{10},\frac{6\sqrt{6}\sigma}{5\hbar}a^\frac{5}{2}\right),
\end{equation}
where $C^J_Y$ are amplitude constants.
These Bessel function are illustrated in the first figure.
Expanding (\ref{sol1}) for small $a$
\begin{eqnarray}
\label{approxsol1}
\psi_1&=&-\frac{2^\frac{3}{10}C_Y}{\Gamma(\frac{7}{10})\sin(\frac{3\pi}{10})}\\
&&+\frac{5}{3}2^\frac{7}{10}\Gamma(\frac{7}{10})
\left(\sin(\frac{3\pi}{10})C_J+\cos(\frac{3\pi}{10})C_Y\right)a^\frac{3}{2}
+{\mathcal O}(a^5)\nonumber \\
&&\approx1.17C_Y+\left(2.84C_J+2.07C_Y\right)a^\frac{3}{2},\nonumber
\end{eqnarray}
so that in particular the limit as $a\rightarrow0$ is given by the finite value of the
first term of (\ref{approxsol1}).
\newline
\begin{figure}
\includegraphics[height=6cm]{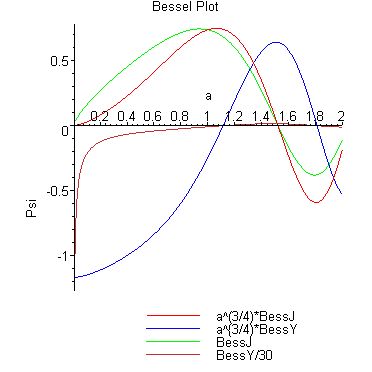}
\caption{Bessel functions for a,  $\sigma=5\hbar/(6\sqrt(6))$}
\label{DiagramTwo}
\end{figure}
\newpage
\section{Relaxation of the scalar field}\label{relaxp}
Relaxing the scalar field in (\ref{clst}) gives line element
\begin{equation}
\label{relaxedscalar}
ds^2=-\sigma^2y\beta(t)^4dt^2+\frac{dy^2}{4y}+\sigma^2y\beta(t)^2d\Sigma_2^2,
\end{equation}
$y$ remains a homothetic Killing potential,  obeying the last two equations of (\ref{hkv}),
regardless of the choice of $\beta$;
$\beta=1/\sqrt{cl}$ recovers the scalar-Einstein solution (\ref{clst}),
this choice of power of $\beta$ is made for later convenience.
After subtracting off the source,  $V_2$ has to vanish or else $y$ is manifest.
The field equations become
\begin{eqnarray}
\label{rsfe}
-\sigma^2y\beta^6{\mathcal G}^t_{.t}&=&\dot{\beta}^2+\beta^4-\sigma^2\beta^6-\beta^2\dot{\xi^2}\\
+\sigma^2y\beta^6{\mathcal G}^\theta_{.\theta}&=&
-\beta\ddot{\beta}+2\dot{\beta}^2+\sigma^2\beta^6-\beta^2\dot{\xi}^2,~~~
{\mathcal G}^r_{.r}=2{\mathcal G}^\theta_{.\theta}+{\mathcal G}^t_{.t}.
\nonumber
\end{eqnarray}
The momenta are
\begin{equation}
\label{rsmom}
\pi_\beta=\frac{2\sigma\dot{\beta}}{\beta^2},~~~
\pi_\xi=-2\sigma\dot{\xi}.
\end{equation}
The Hamiltonian is
\begin{equation}
\label{rsham}
{\mathcal H}_2=\frac{\beta^2}{4\sigma}\pi_\beta^2-\frac{1}{4\sigma}\pi_\xi+U_2
=-\sigma^3y\beta^4{\mathcal G}^t_{.t},~~~~~
U_2=\sigma\beta^2(1-\sigma^2\beta^2),
\end{equation}
and the $\pi_\beta$ Hamilton equation is
\begin{equation}
\label{rspibeta}
\dot{\pi}_\beta+\frac{\partial {\mathcal H}_2}{\partial \beta}=
-2\sigma^3y\beta^3\left({\mathcal G}^\theta_{.\theta}+{\mathcal G}^t_{.t}\right),
\end{equation}
The mini-metric is
\begin{equation}
\label{mini2}
{\mathit M}_{2AB}=\left(
\begin{array}{cc}
\frac{4\sigma}{\beta^2}&0\\
0&-4\sigma
\end{array}
\right),~~
\sqrt{-\det({\mathit M}_2)}=\frac{4\sigma}{\beta}.
\end{equation}
As before using (\ref{quant}) gives the Wheeler-DeWitt equation
\begin{equation}
\label{wdw2}
\frac{4\sigma}{\hbar^2}{\mathcal H}_2\psi
=-\beta\left(\beta\psi_\beta\right)_\beta +\psi_{\xi\xi}
+\frac{4\sigma^2\beta^2}{\hbar^2}\left(1-\sigma^2\beta^2\right)\psi,
\end{equation}
with solution
\begin{equation}
\label{whittakersol}
\psi_2=\frac{1}{\beta}\sum_{+-}A^+_-\exp\left(\frac{\pm\varepsilon\xi}{\hbar}\right)
\sum_{MW}C^M_WW^M_W\left(\frac{\imath}{2\hbar},\frac{\varepsilon}{2\hbar},\frac{2\imath\sigma^2\beta^2}{\hbar}\right),
\end{equation}
where $A_+,A_-,C_M,C_W$ are complex amplitude constants
and $\varepsilon$ is a non-negative real source scalar field constant;
there is a qualitative difference between $\varepsilon=0$ and $\varepsilon\ne0$,
the former jumps at $\beta=0$ the later does not:
only $\psi\psi^\dag$ is measurable and for that there is no jump in either case.
Taking $1=2\sigma=\hbar=2A_+=2A_-=C_M,~0=C_W$ and expanding the WhittakerM function
for small $\beta$
\begin{equation}
\label{expan2}
\psi_W=\frac{1}{2}\left[\beta+\frac{\beta^3}{8}+{\mathcal O}(\beta^5)\right],
\end{equation}
expanding the exponential term for small $\xi$
\begin{equation}
\label{expan3}
\psi_e=\left[1+\frac{\xi^2}{2}+\frac{\xi^4}{24}+{\mathcal O}(\xi^5)\right],
\end{equation}
expanding all (\ref{whittakersol}) to lowest order
\begin{equation}
\psi_2=k'\beta^\varepsilon,
\end{equation}
where $k'$ is a complex constant which varies for different $\varepsilon$.
\begin{figure}
\includegraphics[height=6cm]{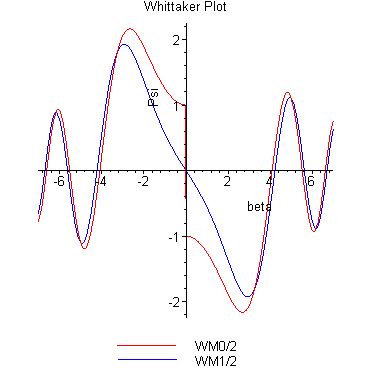}
\caption{$1=2\sigma=\hbar=C_M,~0=C_W$,
red $(\imath -1)W_M(\varepsilon=0)$,
blue $\imath W_M(\varepsilon=1)$.}
\label{DiagramThree}
\end{figure}
\section{Extrapolation,  Approximation and Generalization}\label{eanda}
Extrapolate by combining \S\ref{relaxk} and \S\ref{relaxp}
gives the wavefuncion to lowest order
\begin{equation}
\label{combinyp}
\psi=k'a^\frac{3}{2}\beta^\varepsilon,
\end{equation}
where $k'$ is a complex constant,
transferring to double null coordinates using (\ref{slclprop}) gives
\begin{equation}
\label{combimuv}
\psi=k(uv)^\frac{3}{4}
\left[\frac{r_+r_-}{(v-u)((1-4\sigma^2)v-u)}\right]^\frac{\varepsilon}{2},
\end{equation}
where $k$ is a complex constant.
The singularity is at $u=(1\pm2\sigma)v$ where the wavefunction takes the form
\begin{equation}
\label{wlim}
\psi |_\pm=k(1\pm2\sigma)^\frac{3}{4}v^\frac{3}{2}
\left[\frac{\mp r_\pm}{\sigma(1\pm2\sigma)v}\right]^\frac{\varepsilon}{2},
\end{equation}
substituting for $r_\pm$ for $\varepsilon>0$ the wavefunction vanishes at the singularity:
the desired result.
For $\varepsilon=0$ the wavefunction is a simple function of $v$,
for $\varepsilon<0$ the Whittaker functions (\ref{whittakersol}) are not defined.

There are a several assumptions used in arriving at (\ref{combimuv}).
{\it Firstly} it has been assumed that a wavefunction derived in one segment of the spacetime
can be extended to the whole spacetime,  in particular $v=0$ and $u=0$ regions are not included
in the coordinate systems (\ref{slst}) and (\ref{clst})
and these regions are needed if one wants to study junctions with flat spacetime,
however the curvature singularity exists in both systems in the same sense: in
the double null form (\ref{dnform}) the curvature singularity is at $g^{\theta\theta}$
and the line element truncates here,  similarly for (\ref{slst}) and (\ref{clst}) at $g^{xx}$;
and in this sense the wavefunction exists at the classically singular point.
The Aharanov-Bohm \cite{bi:ES}, \cite{bi:AB}
not only shows the existence of the vector potential it also
shows that the wavefunction is smooth rather than discontinuous at boundaries,
and this justifies the preference of a smooth wavefunction here.
{\it Secondly} no boundary conditions on the quantum system are applied,
these would cut down on the large number of constants in the solutions
(\ref{sol1}) and (\ref{whittakersol}),
for present purposes these are unlikely to make a difference
as we require existence not uniqueness.
{\it Thirdly} no method of extracting information from the wavefunction has been given,
so there is no method of recovering the curvature singularity from it,
it might happen that any such method must itself be in some sense singular,
{\it Fourthly} the wavefunctions in the two regions can be combined and furthermore
done so without considerations of phase.  For large distances the wavefunctions
(\ref{sol1}) and (\ref{whittakersol}) are approximately trigonometric but it is not
clear whether they peak and dip at the same time or not.
The Hamiltonian which is a linear combination of (\ref{ham1}) and
(\ref{rsham}) has a separable solution which is a product of (\ref{sol1}) and
(\ref{whittakersol});  explicitly $4\sigma(H_\beta\psi+\ell H_a\psi)/\hbar^2$ has solution
\begin{equation}
\label{solution}
\psi={\mathcal C}\frac{a^\frac{3}{4}}{\beta}\cosh\left(\frac{\varepsilon\xi}{\hbar}\right)
BJ\left(\frac{3}{10},\frac{6\sqrt{6}\sigma}{5\hbar}a^\frac{5}{2}\right)
WM\left(\frac{\imath}{2\hbar},\frac{\varepsilon}{2\hbar},\frac{2\imath\sigma^2\beta^2}{\hbar}\right),
\end{equation}
where $\mathcal C$ is an amplitude constant,  note (\ref{solution}) is independent of $\ell$.
\section{Conclusion}\label{conc}
The above systems ${\mathcal H}_1,{\mathcal H}_2$ are not restricted to be either exploding
or imploding,  such restrictions might come from additional physical assumptions.
The particle content corresponding to the above wave picture is not clear,
it is not even clear if it at best corresponds to one or many particles:  presumably the content
is of a scalar field so configured that it cancels out the energy of gravitons,
giving no overall energy which would agree with the classical case.
For microscopic application to annihilation and creation operators the above Hamiltonians
${\mathcal H}_1,{\mathcal H}_2$
could be the first step in finding out how spacetime changes.
For macroscopic application to 'black holes' and 'white holes' again the Hamiltonians could
be a first step in solving the 'back reaction' problem.

Our conclusion is that in the specific case studied here
where classical spacetime has curvature singularities
the quantum mechanical wavefunction can be finite,
and that furthermore this is an indication of general behaviour.

\end{document}